# ABSTRACTS

*E.O. Modlo*

**COMPETENCE OF BACHELOR IN ELECTROMECHANICS IN SIMULATION**

The article is devoted to communication competence in modeling with other competences of Bachelor in Electromechanics, its structure and the contribution of components in the formation of competence. The approaches to defining competence of bachelor-electrician are determined. A system of competencies is suggested. In order to determine the inclusion of each of the selected competencies in the formation of competence of Bachelor in Electromechanics, experts' opinions were surveyed. The goals were to determine the structure and content of the bachelor's in Electromechanics competence in modeling. The research focus was to research the relationship of competence in modeling with other competencies of Bachelor of Electromechanics, its structure and the contribution of components in the formation of competence. The object of the research was Bachelor's in Electromechanics learning process. The subject of the research was the theoretical base of expertise of Bachelors in Electromechanics in modeling. The research methods used were: theoretical – analysis of scientific sources concerning the study, classification, specification and synthesis of theoretical, empirical, and experimental data, theoretical design and simulation of competencies of bachelor-electrician in modeling, analysis, synthesis, abstraction, induction, deduction, systematization and classification of results of theoretical research; empirical – the current state of research training in electromechanics, monitoring, summarizing domestic and overseas teaching experience, expert evaluation. The results of the research were the connection of competence in modeling with other competencies of Bachelor in Electromechanics, its structure and the contribution of components in the formation of competence. The structure and content of the electrician's competence in modeling were defined. The principal conclusions and recommendations were: the system of competence in electromechanics in modeling involves three groups of competencies: the general scientific, common professional, and specific professional ones. The formation of competence of Bachelor in Electromechanics in simulation cycle starts in mathematical and natural-scientific training (the leading one is general scientific competence) and continues in the cycle of professional and practical training (the leading ones are common professional and specific professional competencies).

**Є.О. МОДЛО,**
*старший викладач кафедри автоматизованого управління металургійними процесами та електроприводом ДВНЗ «Криворізький національний університет»*

# КОМПЕТЕНТНІСТЬ БАКАЛАВРА ЕЛЕКТРОМЕХАНІКИ В МОДЕЛЮВАННІ

Статтю присвячено питанням зв'язку компетентності в моделюванні з іншими компетентностями бакалавра електромеханіки, її структури та внеску складових у сформованість компетентності. Розглянуто підходи до визначення компетентності бакалавра електромеханіки. **Запропоновано власну систему компетенцій.** З метою визначення внеску кожної з виділених компетенцій у формування компетентності бакалавра електромеханіки проведено опитування експертів.

*Ключові слова: компетентність, система компетенцій, електромеханіка, моделювання.*

**Постановка проблеми у загальному вигляді та її зв'язок з важливими науковими та практичними завданнями.** Однією зі складових системи професійної підготовки сучасного інженера є комп'ютерне моделювання, яке широко використовується за всіма видами інженерної діяльності. Особливої ролі комп'ютерне моделювання набуває у навчанні фахівців галузі знань 0507 «Електротехніка та електромеханіка», забезпечуючи від 60% в циклі математичної, природничо-наукової підготовки до 72% в циклі професійної та практичної підготовки бакалаврів електротехніки та електромеханіки [1]. Це пов'язано з тим, що, з одного боку, математичне моделювання електромеханічних об'єктів та перебігу процесів в електромеханічних системах є одним із видів професійної діяльності інженера-електромеханіка, а з іншого – з тим, що математичне моделювання є основою фундаментальної (фізико-математичної) підготовки інженера-електромеханіка. Тому здатності бакалавра електромеханіки застосовувати методи математичного моделювання, теоретичного та експериментального дослідження із використанням ІКТ є основою загальнопрофесійної компетентності бакалавра електромеханіки в моделюванні.

**Аналіз останніх досліджень та публікацій, в яких започатковано розв'язання проблеми і на які спирається автор.** Професійну підготовку студентів електроенергетичних спеціальностей розглядали М.Г. Дунаєва, О.В. Шищенко, О.Е. Коваленко, О.П. Шаліна, Г.Б. Голубов, С.О. Пчела, О.В. Гамов, О.І. Солошич, Т.В. Крилова. Окремі елементи методики навчання моделювання електромеханічних систем висвітлено у працях О.Д. Тельманової, О.Ф. Бабічевої, К.О. Сороки, О.П. Чорного, О.І. Толочка.

**Виділення невирішених раніше частин загальної проблеми.** Недослідженими залишаються питання зв'язку компетентності в моделюванні з іншими компетентностями бакалавра електромеханіки, її структури та внеску складових у сформованість компетентності.

**Формулювання цілей статті.** Отже, актуальність та недостатній рівень вивчення проблеми дозволили сформулювати *мету* дослідження, яка полягає у визначенні структури та змісту компетентності бакалавра електромеханіки в моделюванні.

**Об'єктом** дослідження є процес навчання бакалаврів електромеханіки.

**Предметом** дослідження є теоретичні засади формування компетентності бакалавра електромеханіки в моделюванні.

Відповідно до поставленої мети були сформульовані такі **завдання**: 1) побудувати систему компетенцій бакалавра електромеханіки в моделюванні; **2) визначити внесок компетенцій у формування компетентності бакалавра електромеханіки в моделюванні.**

З метою розв'язання поставлених завдань застосовувалися теоретичні та емпіричні **методи дослідження**, а саме: теоретичні – аналіз наукових джерел з проблеми дослідження; класифікація, систематизація, конкретизація та узагальнення теоретичних, емпіричних і експериментальних даних; теоретичне проектування та моделювання системи компетенцій бакалавра електромеханіки в моделюванні; аналіз, узагальнення, абстрагування, індукція, дедукція, систематизація і класифікація результатів теоретичного дослідження; емпіричні – дослідження сучасного стану підготовки фахівців з електромеханіки, спостереження, узагальнення вітчизняного та зарубіжного педагогічного досвіду, експертне оцінювання.

**Виклад основного матеріалу дослідження з повним обґрунтуванням отриманих наукових результатів.** *Компетентність бакалавра електромеханіки в моделюванні* визначимо як сформовану в процесі навчання системну властивість особистості, яка містить такі складові: когнітивно-змістову (гносеологічну) – знання; операційно-технологічну (праксеологічну) – навички, уміння, досвід діяльності; ціннісно-мотиваційну (аксіологічну) – мотивація, ціннісне ставлення; соціально-поведінкову – комунікабельність, здатність до адаптації, здатність до інтеграції.

До визначення компетенцій, результатом опанування яких є компетентність бакалавра електромеханіки в моделюванні, існує декілька підходів. Згідно з ОКХ бакалавра електромеханіки [2] виділяють такі групи компетенцій: соціально-особистісні, загальнонаукові, інструментальні, загальнопрофесійні та спеціальні професійні.

Соціально-особистісні компетенції включають в себе розуміння та сприйняття етичних норм поведінки стосовно інших людей і природи (принципи біоетики); розуміння необхідності та дотримання норм здорового способу життя; здатність учитися; здатність до критики й самокритики; креативність, здатність до системного мислення; адаптивність і комунікабельність; наполегливість у досягненні мети; турбота про якість виконуваної роботи; толерантність; екологічна грамотність.

До інструментальних компетенцій відносять здатність до письмової та усної комунікації рідною мовою, знання іншої мови (мов), навички роботи з комп'ютером, навички управління інформацією та дослідницькі навички.

Навчання у ВНЗ без опанування цих компетенцій на достатньому рівні неможливе, тому надалі ці групу компетенцій не розглядатимемо, вважаючи сформованість соціально-особистісної та інструментальної компетентностей майбутнього фахівця з електромеханіки необхідною умовою формування його компетентності в моделюванні.

Таким чином, формування компетентності бакалавра електромеханіки в моделюванні відбувається через систему загальнонаукових, загальнопрофесійних та спеціальних професійних компетенцій. Відповідна складова галузевого стандарту вищої освіти [2] містить посилання на спеціальні професійні компетенції бакалавра електромеханіки, проте не містить їх переліку, тому з метою виділення необхідних компетенцій доцільно звернутися до інших джерел та фахових експертів.

Стандарти підготовки інженерів-електромеханіків, запропоновані Міністерством освіти, навчальними програмами коледжів та університетів Онтаріо (Канада) [3], містять три складові: Vocational standard – аналог спеціальних професійних компетенцій вітчизняного стандарту, Generic employability skills standard – аналог загальнопрофесійних, інструментальних (частково) та загальнонаукових (частково) компетенцій вітчизняного стандарту, та General education standard – аналог соціально-особистісних, інструментальних (частково) та загальнонаукових (частково) компетенцій вітчизняного стандарту.

У результаті опанування *Vocational standard* у випускників має бути сформовано такі компетенції: виготовлення механічних вузлів і агрегатів та збирання електричних компонентів і електронних вузлів, застосовуючи навички і знання основних правил діючих норм техніки безпеки; аналізувати, інтерпретувати і робити електричні, електронні, механічні креслення та складати пов'язані з ними документи і графіки, необхідні для проектування електромеханічних систем; обирати і використовувати випробувальне обладнання та різні методи усунення неполадок для оцінки електромеханічних схем, обладнання, процесів, систем і підсистем; змінювати, підтримувати і ремонтувати електричні, електронні та механічні компоненти обладнання і системи забезпечення їх функціонування відповідно до специфікацій; застосовувати методи інженерії та фундаментальних наук для аналізу і розв'язання проблем проектування та інших складних технічних завдань для виконання робіт, пов'язаних з електромеханічною інженерією; розробляти та аналізувати механічні компоненти, процеси та системи на основі застосування інженерних принципів і рішень; застосовувати принципи механіки твердого тіла та механіки рідин для розробки та аналізу електромеханічних систем; проектувати, аналізувати, будувати і виправляти логічні та цифрові схеми, пасивні та активні схеми постійного і змінного струму; проектувати, обирати, застосовувати, інтегрувати і усувати неполадки в різних промислових системах управління електроприводом, пристроях збору даних та системах; проектувати, аналізувати і шукати несправності у мікропроцесорних системах; встановлювати та налаштовувати комп'ютерне обладнання і мови програмування високого рівня для підтримки електромеханічного інженерного середовища; аналізувати, програмувати, встановлювати, інтегрувати та налаштовувати автоматизовані системи, у тому числі робототехнічні; виконувати та підтримувати системи інвентаризації, запису та документування; допомагати в управлінні проектами, застосовуючи принципи ведення бізнесу до електромеханічного інженерного середовища; обирати електромеханічне обладнання, компоненти та системи, що відповідають робочим вимогам і функціональним специфікаціям; складати, координувати і реалізувати програми і процедури контролю якості і забезпечення якості; виконувати всі роботи відповідно до законів, правил, кодексів, настанов та процедур забезпечення безпеки; розвивати особистісні та професійні стратегії і плани підвищення продуктивності праці і робочих відносин з клієнтами, колегами і керівниками [3, 6–7].

*Generic employability skills standard* визначає такі компетенції: спілкуватися ясно, коротко і правильно у письмовій, усній і візуальній формі відповідно до мети та потреб аудиторії; переосмислювати інформацію, ідеї та концепції за допомогою усного, візуального, числового та символьного подання, які демонструють їх розуміння; застосовувати широкий спектр математичних методів із ступенем точності, необхідним для вирішення проблеми і прийняття рішення; доречно використовувати різне комп'ютерне обладнання, програмне забезпечення та інші технічні засоби відповідно до виконаних завдань; взаємодіяти з колегами та іншими групами так, щоб сприяти ефективним робочим відносинам і досягненню цілей; саморефлексувати всі кроки і процеси, що використовуються у вирішенні проблем та прийнятті рішення; збирати, аналізувати та упорядковувати доцільну та необхідну інформацію з різних джерел; оцінювати обґрунтованість аргументів, базованих на якісній та кількісній інформації для того, щоб прийняти або піддати сумніву висновки, зроблені іншими; створювати інноваційні стратегії та/або продукти, які відповідають виявленим потребам; управляти використанням часу та інших ресурсів для досягнення особистих цілей та/або цілей проекту; брати на себе відповідальність за власні дії та рішення; адаптуватися до нових ситуацій і вимог шляхом застосування та/або оновлення власних знань і навичок; реалістично репрезентувати власні навички, знання і досвід в особистих цілях та для забезпечення продуктивної зайнятості [3, 27].

Останній блок – *General education standard* – визначає ключові компетенції: естетичні (розуміти красу, форму, смак і роль мистецтва в житті суспільства), громадянські (розуміти сенс свободи, права, брати участь у громадському та суспільному житті), міжкультурні (розуміти культурне, соціальне, етнічне та мовне розмаїття країни та світу), саморозвитку (прагнути до більшої самосвідомості, інтелектуального зростання, благополуччя та розуміння інших), соціальні (розуміти відносини між людьми і суспільством), загальнонаукові (оцінювати внесок науки в розвиток цивілізації, людське розуміння та потенціал), загальнотехнологічні (розуміти взаємозв'язок між розвитком та використанням технологій, суспільством та екосистемою), економічні (розуміти сенс, історію та організацію праці, а також зміни у робочому житті людей і суспільства) [3, 46–48].

Фахівці Human Resource Systems Group [4] визначають дві групи компетенцій, що можуть бути сформовані на одному з п'яти рівнів (1 – базовий, 5 – експертний):

1) *загальні* компетенції, до складу яких входять компетенції: з письмової комунікації (на рівні 4 – передавання складної інформації); аналітичного мислення (на рівні 4 – засто-сування загального аналізу); усної комунікації (на рівні 4 – передавання складних повідомлень); розв'язання проблем (на рівні 4 – розв'язання складних проблем); планування та організації (на рівні 4 – планування та організація багатокомпонентної складної діяльності); лідерства у команді (на рівні 3 – формування сильних команд); критичного мислення (на рівні 4 – формулювання загальних стратегій з багатовимірних стратегічних питань); прогнозування та управління (на рівні 3 – автоматичне управління та оперативна підтримка);

2) *технічні* компетенції, до складу яких входять компетенції: з калібрування/математики (на рівні 4 – багатокрокові складні обчислення); інструментально-технологічні (на рівні 4 – експлуатація ручних та електроінструментів усіх типів); проектування (на рівні 4 – демонстрація додаткових знань і здібностей, застосування компетенції у нових або складних ситуаціях, консультування інших фахівців); обслуговування та відновлення електричних систем (на рівні 4 – демонстрація додаткових знань і здібностей, застосування компетенції у нових або складних ситуаціях, консультування інших фахівців); електричної та електронної інженерії (на рівні 5 – демонстрація експертних знань та здатностей, застосування компетенції у найбільш складних ситуаціях, розробка нових підходів та методів, визнання як експерта у співтоваристві та/або за його межами); роботи із електричним обладнанням (на рівні 5 – експертному).

Матриця В компетенцій для сектора електронної та електричної інженерії [5, 14–15], розроблена у рамках проекту європейського VQTS II (Vocational Qualification Transfer System), охоплює 8 груп компетенцій, кожна з яких визначена на 3 або 4 рівнях: 1) **підготовка**, планування, монтаж і установка електричних та/або електронних систем для будівельного та промислового застосування; 2) перевірка, догляд та обслуговування електричних та/або електронних систем і устаткування; 3) створення, введення в експлуатацію і регулювання електричних та/або електронних систем; 4) проектування, модифікація та адаптація провідних мереж та друкованих плат для електричних та/або електронних систем, включаючи їх інтерфейси; 5) розробка спеціалізованих електричних та/або електронних проектів; 6) нагляд і підтримка виробничих і бізнес-процесів, включаючи управління якістю; 7) установка, налагодження і тестування модифікації прикладного програмного забезпечення для установки та експлуатації електричних та/або електронних систем; 8) діагностика та ремонт електричних/електронних систем і устаткування.

Проведений аналіз різних підходів до визначення компетенцій бакалавра електромеханіки дозволив запропонувати власну систему компетенцій (рис. 1).

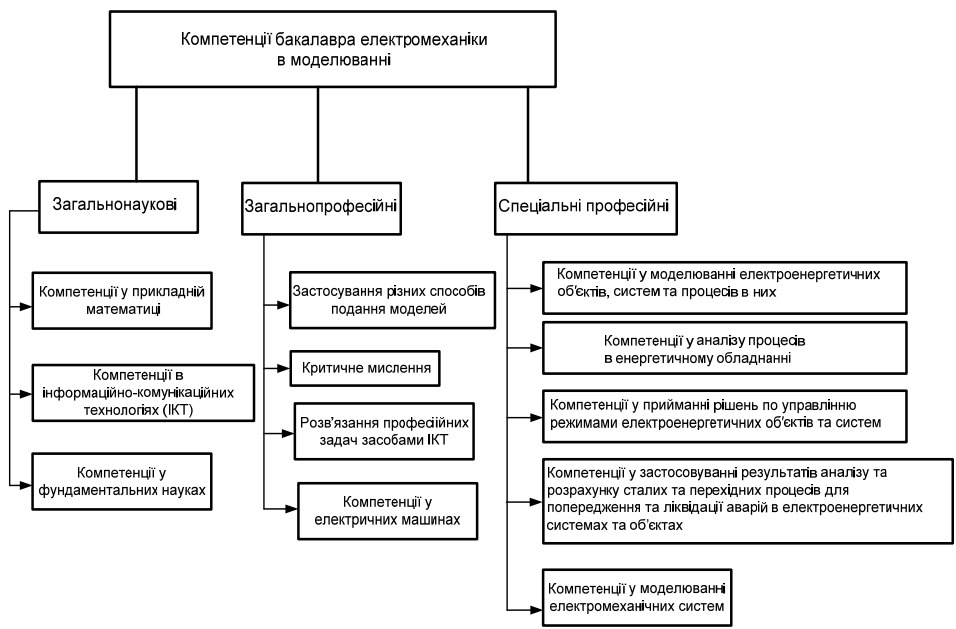

**Рис. 1. Система компетенцій бакалавра електромеханіки в моделюванні**

З метою визначення внеску кожної з виділених компетенцій у формування компетентності бакалавра електромеханіки в моделюванні у квітні 2013 р. було проведено опитування експертів – провідних фахівців із підготовки бакалаврів електромеханіки.

Загальна кількість учасників опитування – 14 осіб, з яких 7% – доктори наук, 43% – кандидати наук, 50% – старші викладачі випускових кафедр. 85% опитаних мають значний досвід упровадження електромеханічних систем у виробничий процес підприємств гірничо-металургійного комплексу.

Опрацювання відповідей на перше питання опитування («Розташуйте групи компетенцій за значущістю (0 – найвища, 2 – найнижча): а) загальнонаукові; б) загальнопрофесійні; в) спеціальні професійні»), спрямоване на визначення внеску кожної групи компетенцій у сформованість компетентності бакалавра електромеханіки в моделюванні, показало, що внесок загальнопрофесійних та спеціальних професійних компетенцій однаковий – по 35,3% кожна група, у той час як внесок загальнонаукових – 29,4%.

У групі загальнонаукових компетенцій провідне місце займають компетенції у прикладній математиці (рис. 2а), а серед загальнопрофесійних – критичне мислення (рис. 2б).

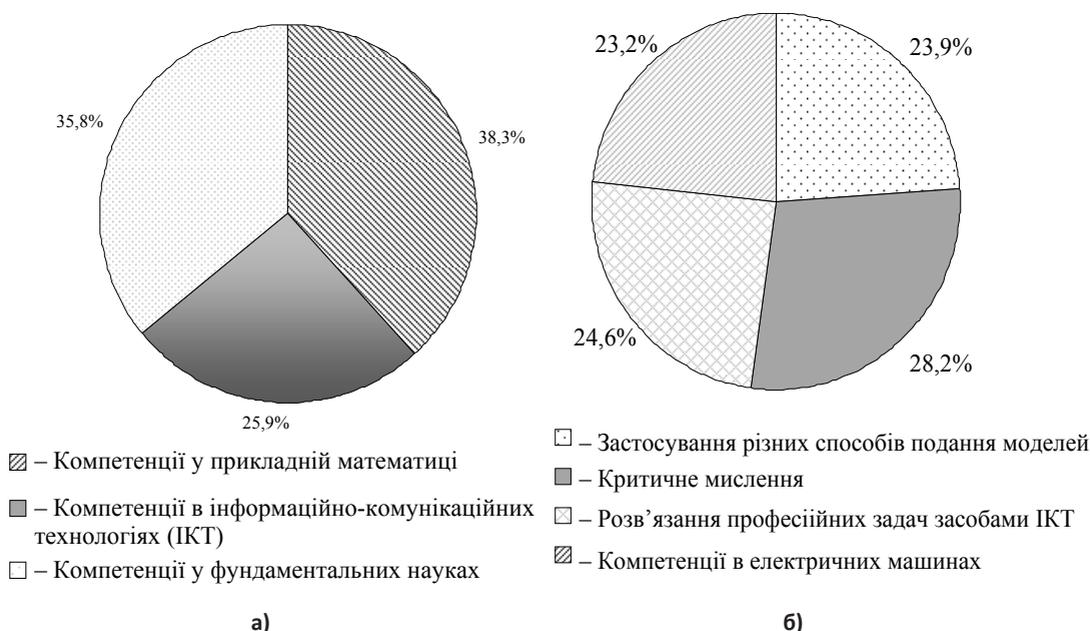

▨ – Компетенції у прикладній математиці
■ – Компетенції в інформаційно-комунікаційних технологіях (ІКТ)
□ – Компетенції у фундаментальних науках

⋅⋅ – Застосування різних способів подання моделей
■ – Критичне мислення
▥ – Розв'язання професійних задач засобами ІКТ
▨ – Компетенції в електричних машинах

**а)**      **б)**
**Рис. 2. Внесок загальнонаукових (а) та загальнопрофесійних (б) компетенцій
у формування компетентності бакалавра електромеханіки в моделюванні**

Усі експерти погоджуються з тим, що серед спеціальних професійних компетенцій найбільш вагомими (24,7%) для формування компетентності бакалавра електромеханіки в моделюванні є компетенції у моделюванні електромеханічних систем, у той час як вагомість кожної із споріднених компетенцій інженера-електроенергетика становить у середньому 18,8% (рис. 3).

Результати опитування дали можливість визначити внесок кожної компетенції у систему компетенцій бакалавра електромеханіки в моделюванні (табл. 1).

*Таблиця 1*
**Внесок компетенцій у формування компетентності бакалавра електромеханіки в моделюванні**

| Компетенція | Внесок |
|---|---|
| *Загальнонаукові, у т. ч.* | |
| компетенції у прикладній математиці | 38,3% |
| компетенції в інформаційно-комунікаційних технологіях (ІКТ) | 25,9% |

| Компетенція | Внесок |
|---|---|
| компетенції у фундаментальних науках | 35,8% |
| *Загальнопрофесійні, у т. ч.* | |
| застосування різних способів подання моделей | 23,9% |
| критичне мислення | 28,2% |
| розв'язання професійних задач засобами ІКТ | 24,6% |
| компетенції в електричних машинах | 23,2% |
| *Спеціальні професійні, у т. ч.* | |
| компетенції в моделюванні електроенергетичних об'єктів, систем та процесів у них | 19,5% |
| компетенції в аналізі процесів в енергетичному обладнанні | 18,1% |
| компетенції у прийманні рішень з управління режимами електроенергетичних об'єктів та систем | 20,0% |
| компетенції у застосовуванні результатів аналізу та розрахунку сталих та перехідних процесів для запобігання та ліквідації аварій в електроенергетичних системах та об'єктах | 17,7% |
| компетенції у моделюванні електромеханічних систем | 24,7% |

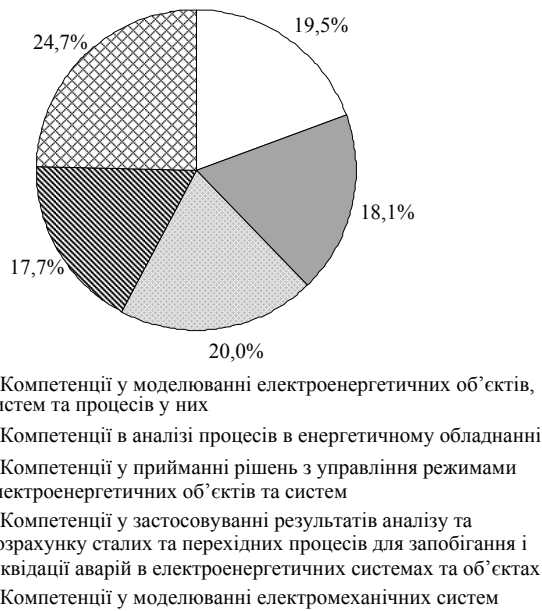

☐ – Компетенції у моделюванні електроенергетичних об'єктів, систем та процесів у них
▨ – Компетенції в аналізі процесів в енергетичному обладнанні
▦ – Компетенції у прийманні рішень з управління режимами електроенергетичних об'єктів та систем
▧ – Компетенції у застосовуванні результатів аналізу та розрахунку сталих та перехідних процесів для запобігання і ліквідації аварій в електроенергетичних системах та об'єктах
▨ – Компетенції у моделюванні електромеханічних систем

**Рис. 3. Внесок спеціальних професійних компетенцій у формування компетентності бакалавра електромеханіки в моделюванні**

**Висновки за результатами дослідження, перспективи подальших розвідок в означеному напрямі.**

1. Компетентність бакалавра електромеханіки в моделюванні – це сформована в процесі навчання системна властивість особистості, яка містить такі складові: когнітивно-змістову (гносеологічну) – знання; операційно-технологічну (праксеологічну) – навички, уміння, досвід діяльності; ціннісно-мотиваційну (аксіологічну) – мотивація, ціннісне ставлення; соціально-поведінкову – комунікабельність, здатність до адаптації, здатність до інтеграції.

2. Система компетенцій бакалавра електромеханіки в моделюванні включає в себе три групи компетенцій: загальнонаукові (у прикладній математиці; в інформаційно-комунікаційних технологіях; у фундаментальних науках), загальнопрофесійні (застосування різних способів подання моделей; критичне мислення; розв'язання професійних задач засобами ІКТ; у електричних машинах), спеціальні професійні (у моделюванні електроенергетичних об'єктів, систем та процесів в них; в аналізі процесів в енергетичному обладнанні; у прийманні рішень з управління режимами електроенергетичних об'єктів та систем; у застосовуванні результатів аналізу та розрахунку сталих і перехідних процесів для запобігання та ліквідації аварій в електроенергетичних системах та об'єктах; у моделюванні електромеханічних систем).

3. Відповідно до виділеної у дослідженні структури підготовки з моделювання бакалаврів електромеханіки, формування компетентності бакалавра електромеханіки в моделюванні розпочинається у циклі математичної та природничо-наукової підготовки (провідними є загальнонаукові компетенції) і продовжується у циклі професійної та практичної підготовки (провідними є загальнопрофесійні та спеціальні професійні компетенції). Враховуючи, що найвищий рівень системності у процесі її формування досягається при підготовці до державної атестації (іспит, захист кваліфікаційної роботи), доцільним є при оцінюванні рівня сформованості ураховувати також її виявлення студентами під час державної атестації.

Статья посвящена вопросам связи компетентности в моделировании с другими компетентностями бакалавра электромеханики, ее структуры и вклада составляющих в сформированность компетентности. Рассмотрены подходы к определению компетентности бакалавра электромеханики. Предложена собственная система компетенций. С целью определения вклада каждой из выделенных компетенций в формирование компетентности бакалавра электромеханики проведен опрос экспертов.

*Ключевые слова: компетентность, система компетенций, электромеханика, моделирования.*

The article is devoted to communication competence in modeling with other competences of Bachelor in Electromechanics, its structure and the contribution of components in the formation of competence. The approaches to defining competence of bachelor-electrician are determined. A system of competencies is suggested. In order to determine the inclusion of each of the selected competencies in the formation of competence of Bachelor in Electromechanics, experts' opinions were surveyed. The goals were to determine the structure and content of the bachelor's in Electromechanics competence in modeling. The formation of competence of Bachelor in Electromechanics in simulation cycle starts in mathematical and natural-scientific training (the leading one is general scientific competence) and continues in the cycle of professional and practical training (the leading ones are common professional and specific professional competencies).

*Key words: competence, competency system, electromechanics, modeling.*